\documentclass[11pt,a4paper]{article}
\pdfoutput=1

\usepackage{jcappub}

\usepackage[utf8]{inputenc}
\usepackage{url}
\providecommand{\e}[1]{\ensuremath{\times 10^{#1}}}

\title{Generic slow-roll and non-gaussianity parameters in $f(R)$ theories}

\author[a,b]{Tays Miranda}
\author[c,d]{Celia Escamilla-Rivera,}
\author[e,f]{Oliver F. Piattella}
\author[e,g]{and J\'ulio C. Fabris}

\affiliation[a]{Department of Physics, Universidade Federal do Esp\'irito Santo, Avenida Fernando Ferrari 514, zip 29075-910, Vit\'oria, Esp\'irito Santo, Brazil.}
\affiliation[b]{Institute of Cosmology $\&$ Gravitation, University of Portsmouth, Dennis Sciama Building, Burnaby Road, Portsmouth, PO1 3FX, United Kingdom.}
\affiliation[c]{Instituto de Ciencias Nucleares, Universidad Nacional Aut\'onoma de M\'exico, Circuito Exterior C.U., A.P. 70-543, M\'exico D.F. 04510, M\'exico.}
\affiliation[d] {ICTP, Strada Costiera 11, 34151 Trieste, Italy.}
\affiliation[e]{N\'{u}cleo Cosmo-ufes $\&$ Departamento de F\'isica, Universidade Federal do Esp\'irito Santo, Avenida Fernando Ferrari 514, zip 29075-910, Vit\'oria, Esp\'irito Santo, Brazil.}
\affiliation[f]{Institut f\"ur Theoretische Physik, Ruprecht-Karls-Universit\"at Heidelberg, \\ Philosophenweg 16, 69120 Heidelberg, Germany.}
\affiliation[g]{National Research Nuclear University MEPhI, Kashirskoe sh. 31, Moscow 115409, Russia.}

\abstract{
In this paper we establish \textit{formulae} for the inflationary slow-roll parameters $\epsilon$, $\eta$ and $\zeta$ as functions of the Ricci scalar $R$ for $f(R)$ theories of gravity. As examples, we present the analytic and  numerical  solutions of $\epsilon$, $\eta$ and $\zeta$ as functions of the number of e-folds $N$ in two important instances: for the Starobinsky model and for a $f(R)$ reconstruction of the $\alpha$-Attractors. The highlight of our
proposal is to rewrite the slow-roll parameters in terms of $f(R)$, which allows to find directly $n_{\rm s}$, $r$, $\alpha_{\rm s}$ and
$f^{\text{equil}}_{\text{NL}}$ as functions of $R$ itself.
We obtain that both models indicate a small contribution to the non-Gaussianity parameters, which are in good agreement with current observational constraints.}

\begin{document}
\maketitle

\flushbottom


\section{Introduction}
Cosmology has entered an era of relevant progress in high-precision observation.
In this landscape the determinations of the cosmological parameters has led to a clear picture of the history of our universe that seems to favor the inflationary scenario \cite{Starobinsky:1980te, Guth:1980zm, Linde:1981mu}. In this context, departures from the statistical isotropy of the perturbations \cite{Erickcek:2008jp}, from their adiabaticity \cite{Gordon:2000hv} and from Gaussian initial conditions \cite{DeFelice:2010aj,Nojiri:2017ncd} are under scrutiny in order to test the inflationary paradigm, being that any observed deviation from these conditions would open new studies related to the inflationary scenario.

Furthermore, a future detection of large non-Gaussianities in the Cosmic Microwave Background (CMB) would falsify the simplest inflationary scenario, namely the single-field slow-roll inflation. Thanks to CMB observations performed by the WMAP satellite \cite{WMAP:2013} via its information of the scalar spectral index $n_s$ as well as the tensor-to-scalar ratio $r$, some of these models have been already ruled out. While the observables $n_s$ and $r$ are derived in linear cosmological perturbation theory, it is also possible to distinguish among inflationary models by comparing non-Gaussianity of primordial perturbations with the Planck survey data \cite{Ade:2015lrj}, where the canonical single field slow-roll inflation can only give rise to negligible amount of non-Gaussianity. 

Along these lines of thought, effort has been made in exploring the production of large non-Gaussianities in single and multiple-fields inflationary models \cite{Escamilla-Rivera:2015ova}. For single-field inflation, large non-Gaussianities are easily produced in models with a small speed of sound. But for multi-field models, the non-Gaussianities can still be larger so computing them is an important step towards understanding N-flation.

In these regards, first quantitative studies about primordial power spectra of scalar and tensor perturbations for an inflationary model have been presented in \cite{Starobinsky:1983zz}. Furthermore, $f(R)$ models have been considered in order to compute the spectrum of primordial scalar perturbations generated in this inflationary stage \cite{Hwang:1996xh,Liu:2018hno}.
On a parallel set of developments, there has been recent interest in developing an analogy at high curvature $R$ between $f(R)$ models and the $\alpha$-Attractors \cite{Odintsov:2016vzz, Miranda:2017}. 
Inspired by these analysis, in this work we propose a mapping between the slow-roll parameters and non-Gaussianities using a generic form of $f(R)$ in order to find an easy way to add these results as a pipeline over the constraining of the observational parameters. By generic form of $f(R)$ we mean that in principle we could obtain the slow-roll parameters for any $f(R)$, since to apply our definitions we just need a $f(R)$ model and its derivatives in terms of $R$. On the other hand, we need to stress that slow-roll inflation may only occur for the range of $R$ in which $f(R)/R^{2}$ is a slowly changing function of $R$, namely, where its first and second derivatives with respect to $\ln R$ are small by modulus, as shown in \cite{Appleby:2009uf}.

\section{Inflationary background in $f(R)$}\label{Sec:InflBack}
We start with the usual $f(R)$ action given by:
\begin{equation}
S = \frac{M^2_{\rm Pl}}{2} \int{\sqrt{-g}f(R) d^4x}\;,
\end{equation}
where $M_{\rm Pl}$ is the Planck mass and $f(R)$ is an arbitrary smooth function of the Ricci scalar $R$. The field equations, associated to this action,
are given by:
\begin{eqnarray}\label{eq:mov1}
R_{\mu\nu} f^\prime(R) -\frac{1}{2} g_{\mu\nu}f(R) -\nabla_{\mu}\nabla_{\nu} f^\prime (R) 
+ g_{\mu\nu} \Box f^\prime (R) = 0,\; \quad
\end{eqnarray}
where $f^\prime =\partial_{R} f$. 
The above field equations can be recast as:
\begin{eqnarray}
f^\prime (R)G_{\mu\nu} &=& g_{\mu\nu} \frac{f(R)-f^\prime (R)R}{2} + \nabla_{\mu}\nabla_{\nu} f^\prime (R) 
\nonumber\\ && -g_{\mu\nu}\Box f^\prime (R)\;,
\end{eqnarray}
and read effectively as the usual Einstein equations $G_{\mu\nu} = \kappa T^{\text{eff}}_{\mu\nu}$ if we define:
\begin{eqnarray}
T^{\text{eff}}_{\mu\nu} = \frac{1}{\kappa f^\prime} \left[\frac{g_{\mu\nu}}{2} (f - f^\prime R) +  (\nabla_{\mu}\nabla_{\nu}-g_{\mu\nu}\Box)f^\prime\right]\;,
\end{eqnarray}
as an effective energy-momentum tensor. From now on we drop the explicit dependence of $f(R)$ on $R$ in order to keep a lighter notation.

Also we will consider a homogenous, isotropic universe described by a flat metric: 
\begin{equation}\label{eq:flatmetric}
ds^2 = -dt^2 + \delta_{ij}dx^idx^j\;.
\end{equation} 
Under these assumptions we vary the action to obtain the following modified Friedmann equations
\begin{align}\label{modFriedeq}
&3H^2 =\frac{1}{f^\prime} \left[\frac{Rf^\prime -f}{2} -3H\dot{R}f^{\prime\prime}\right]\;, \\
&2\dot{H} +3H^2 =-\frac{1}{f^{\prime}} \left[\frac{f - Rf^\prime}{2}  + 2H\dot{R}f^{\prime\prime} 
+\ddot{R}f^{\prime\prime} +\dot{R}^2 f^{\prime\prime\prime}\right],\; \quad
\end{align}
where the dot denotes derivation with respect to the cosmic time $t$.
Also, we have from the Ricci scalar definition that:
\begin{eqnarray}\label{hdot}
\dot{H} = \frac{R}{6} - 2H^2\;.
\end{eqnarray}

A second-order differential equation for $H$ can be obtained by substituting Eq.~(\ref{hdot}) in the modified Friedmann equation Eq.\eqref{modFriedeq}:
\begin{eqnarray}\label{eq:friedmanngen}
\ddot{H} + \dot{H}(4H-\lambda) -\lambda H^2 +\frac{f}{36Hf''} = 0\;,
\end{eqnarray}
where we have defined $\lambda \equiv f'(6Hf'')^{-1}$. 
This latter has been already analysed for the case  $f(R) = R + \alpha R^{2}$ in \cite{Starobinsky:1980te, DeFelice:2010aj}. Also, it has to be noted that equations for a spatially flat universe given by Eq.(\ref{eq:flatmetric}) can be reduced to one first-order differential equation, for instance see \cite{Motohashi:2017vdc}.

As an illustrative example on how we can recover the standard inflation scenario using Eq.(\ref{eq:friedmanngen}), let us consider an exponential $f(R)$ model, due to its simplicity, of the form 
\begin{eqnarray}\label{eq:expfR}
f(R)=e^{\xi R}\;.
\end{eqnarray}
This choice can be justified by the fact that in the low energy limit $\xi R\ll1$ we get $f(R)=1 + \xi R$, which recovers General Relativity (GR) with an additional constant term. See e.g. \cite{Cognola:2007zu, Elizalde:2010ts, Bamba:2014jia} as pioneering works on exponential $f(R)$ in inflation.
Now, using Eq.~(\ref{eq:expfR}) in Eq.~(\ref{eq:friedmanngen}) we obtain the modified evolution equation
\begin{eqnarray}\label{eq:diff}
\ddot{H} +\dot{H} \left[4H -\frac{1}{6\xi H}\right] -\frac{H}{6\xi} +\frac{1}{36\xi^2 H} = 0\;.
\end{eqnarray}
In the inflationary regime $H \approx$ constant, therefore we can propose a solution of the form
\begin{equation}
H=\bar{H} +\dot{H}\Delta t +\frac{\ddot{H}}{2} (\Delta t)^2 +\ldots \;
\end{equation}
where $\bar{H}^2 =1/6\xi\;.$
With this solution, it is straightforward to calculate the usual slow-roll parameters 
\begin{equation}
\epsilon=-\frac{\dot{H}}{\bar{H}^2}, \quad \dot{H}=-\bar{H}\epsilon, \quad \eta=\frac{\dot{\epsilon}}{\bar{H}\epsilon}\;.
\end{equation}
The latter equations show how to recover the inflationary standard case with our toy model given by Eq.(\ref{eq:expfR}). Nevertheless, metastable slow-roll inflation with a large number of e-folds may not occur at all for this exponential function, since it violates the condition that $f(R)/R^{2}$ has to be almost constant. Furthermore, even if this kind of modification has as initial motivation to describe unified inflation and Dark Energy expansions, they might have singularity problems as discussed in \cite{Appleby:2009uf}.


\section{Generic slow-roll parameters for a $f(R)$ model.}
After having understood how we can recover the usual GR slow roll parameters, 
a generic $f(R)$ theory can be mapped via a conformal transformation into GR with a matter content constituted by a scalar field $\chi$, defined in the following way:
\begin{equation}
\chi \equiv \sqrt{\frac{3}{2}}M_{\rm Pl}\ln f'\;,
\end{equation}
provided that $f'>0$, which is one of the conditions for a viable $f$ theory (it corresponds to an attractive gravity, instead of repulsive, and avoids thus Ostrogradsky instability). This scalar field $\chi$ is subject to the following potential:
\begin{equation}\label{Upotential}
U \equiv \frac{M_{\rm Pl}^2}{2f'^2}(Rf' - f)\;.
\end{equation}
To calculate the slow roll parameters, let us consider the definition
\begin{eqnarray}\label{eq:SR}
\epsilon &\equiv& \frac{M_{\rm Pl}^2}{2}\left(\frac{U_\chi}{U}\right)^2\;, \\
\eta &\equiv& M_{\rm Pl}^2\frac{U_{\chi\chi}}{U}\;, \\
\zeta &\equiv& M_{\rm Pl}^4\frac{U_{\chi}U_{\chi\chi\chi}}{U^2}\;,
\end{eqnarray}
where
\begin{equation}
U_\chi \equiv \frac{dU}{d\chi} = \frac{dU/dR}{d\chi/dR}\;.
\end{equation}
We can employ the definition of the field $\chi$ and its potential $U$ in order to find:
\begin{eqnarray}\label{eq:epsilong}
\epsilon &=& \frac{1}{3}\left(\frac{2f - Rf'}{Rf' - f}\right)^2\;. \\
\eta &=& \frac{2(f')^2}{3f''(Rf' - f)} - \frac{2Rf'}{Rf' - f} + \frac{8}{3}\;,\label{eq:etag} \\
\zeta &=& \frac{4}{9} \frac{\left(2f - Rf'\right)}{\left(Rf' - f\right)^2}\left[-\frac{f''' (f')^{3}}{(f'')^{3}} - \frac{3 (f')^{2}}{f''} + 8f- Rf'\right].\;\label{eq:zetag} \quad
\end{eqnarray}
Applying these generic formulae 
to the Starobinsky model $f = R + \alpha R^2$, we can easily compute:
\begin{align}
&\epsilon = \frac{1}{3\alpha^2 R^2}\;,\\
&\eta = \frac{1-2 \alpha  R}{3 \alpha ^2 R^2}\;,\\
&\zeta =\frac{2(2\alpha R -3)}{9\alpha^3 R^3}\;.
\end{align}
It is straightforward to notice that the above slow-roll parameters tend to zero for $\alpha R\to \infty$, as we expect from the fact that the potential $U$ defined in Eq.~\eqref{Upotential} presents a plateau for the Starobinsky model for $\alpha R\to \infty$. 

It is remarkable that \textit{we are able to express the slow-roll parameters as functions of $R$ for any $f(R)$ theory}. However, in order to make contact with observation, we need to relate $R$ to the number of e-folds $N$ from the end of inflation. We know that the total number of inflationary e-folds should exceed about $60$ in order to solve the horizon and flatness problems, that is:
\begin{equation}
N_{\text{total}} =\ln \frac{a_{\text{end}}}{a_{\text{start}}} \geq 60\;.
\end{equation}
The precise value depends on the energy scale of inflation and on the details of reheating after inflation. The fluctuations observed
in the CMB are created during approximately $N_{\text{CMB}}\approx 50-60$ e-foldings before the end of inflation. Therefore, we use
these numbers to have a precise prediction on the values of the slow-roll parameters and then on the spectral index. 

A simple way to compute $N$ for a generic $f(R)$ theory in the slow-roll approximation is the following:
\begin{eqnarray}\label{eq:Nefolding}
N &\approx& \frac{1}{M_{\rm Pl}}\int_{\chi_1}^{\chi_2}d\chi\frac{1}{\sqrt{2\epsilon}} \nonumber\\
&=& \frac{3}{2}\int_{R_f}^{R_i}dR\left(\frac{f''}{f'}\right)\frac{Rf' - f}{2f - Rf'}\;,
\end{eqnarray}
where the subscripts $i$ and $f$ over the integral denote an initial arbitrary moment during the inflationary phase and the final period of inflation, given by the condition $\epsilon_f \approx 1$.

In order to perform the integration in Eq.(\ref{eq:Nefolding}), we need an explicit form for our $f(R)$ theory. Again, for the Starobinsky model we get:
\begin{equation}\label{eq:approxStar}
N \approx 3\int_{R_f}^{R_i}dR\frac{\alpha^2 R}{1 + 2\alpha R} \sim \frac{3}{2}\alpha(R_i - R_f)\;,
\end{equation}
where we are assuming $\alpha R \gg 1$, which is a necessary condition to have an inflation stage. Considering $R_i \gg R_f$ and dropping the subscript $i$, we can write:
\begin{equation}
\alpha R \sim \frac{2N}{3}\;.
\end{equation}
Now we are able to write the slow-roll parameters as:
\begin{equation}
\epsilon = \frac{3}{4N^2}\;, \qquad \eta = -\frac{1}{N}\;, \qquad \zeta = \frac{1}{N^{2}}\;,
\end{equation}
and the tensor-to-scalar ratio and scalar spectral index as:
\begin{equation}\label{rnsstar}
r \equiv 16\epsilon = \frac{12}{N^2}\;, \qquad n_{\rm s} \equiv 2\eta +1= 1 - \frac{2}{N}\;,
\end{equation}
which for $N = 50-60$ give predictions in very good agreement with observations.

The running of the spectral index can be written as:
\begin{eqnarray}
\alpha_{\rm s} \equiv \frac{d n_{\rm s}}{d\ln k} = -2\zeta + 16\eta\epsilon - 24\epsilon^{2} = -\frac{2}{N^{2}}\;,
\end{eqnarray}
which it is of the same order of the tensor-to-scalar ratio.

As a second case, we consider the $\beta$-model \cite{Miranda:2017}, which is a $f(R)$ approximated reconstruction of the $\alpha$-Attractors:
\begin{eqnarray}
f(R)=R+\sigma^{\beta}R^{\beta+1} +\frac{\sigma}{2}R^2,
\end{eqnarray}
where $\sigma=(1-\beta)^2/3M^2$. We notice that 
when $\beta = 0$ we recover the Starobinsky model modified by a constant in the linear term and when $\beta = 1$ we recover GR.

Using Eqs.~(\ref{eq:epsilong})-(\ref{eq:etag})-(\ref{eq:zetag}) we can compute the following slow-roll parameters and the number of e-folds for this model:
\begin{eqnarray}
&&\epsilon = \frac{4}{3}\left[\frac{1+(1-\beta)(\sigma R)^{\beta}}{2\beta (\sigma R)^\beta +\sigma R}\right]^{2}\;,\\
&&\eta =\frac{8}{3} - 4\left[\frac{1+(\beta+1)(\sigma R)^\beta +\sigma R}{2\beta (\sigma R)^\beta +\sigma R}\right] \\&&
+ \frac{4}{3}\left\{\frac{1+2(\beta +1)(\sigma R)^{\beta}[1+\sigma R]+(\beta + 1)^{2}(\sigma R)^{2\beta}+2\sigma R+(\sigma R)^{2}}{\sigma R \left[\beta (\beta +3)(\sigma R)^\beta +\sigma R +2\beta^2 (\beta+1)(\sigma R)^{2\beta-1}\right]}\right\}\;,\nonumber\\
&&\zeta = \frac{16}{9}\frac{\left[1+(1-\beta)(\sigma R)^{\beta}\right]}{\left[2\beta(\sigma R)^{\beta}+\sigma R\right]^{2}}\Bigg\{-\frac{\beta (\beta + 1)(\beta - 1)\left[1+(\beta +1)(\sigma R)^{\beta}+\sigma R\right]^{3}(\sigma R)^{\beta}}{(\sigma R)^{3}\left[1+\beta (\beta + 1)(\sigma R)^{\beta -1}\right]^{3}}\nonumber\\&&
-3\frac{\left[1+(\beta + 1)(\sigma R)^{\beta}+ \sigma R\right]^{2}}{\sigma R\left[1+\beta (\beta + 1)(\sigma R)^{\beta -1}\right]} +7+ (7 - \beta) (\sigma R)^{\beta}+ 3 \sigma R\Bigg\}\;,\\
&&\label{efoldseq} N = \frac{3}{4}\int d(\sigma R)\left[\frac{2\beta^{2}(\beta +1)(\sigma R)^{2\beta-1}+\beta (\beta+3)(\sigma R)^{\beta}+\sigma R}{1+2(\sigma R)^{\beta}+ (\beta +1)(1-\beta)(\sigma R)^{2\beta}+(1-\beta)(\sigma R)^{\beta +1}+\sigma R}\right].
\end{eqnarray}
As it was shown in \cite{Miranda:2017}, the viable range for $\beta$, determined from the Planck 2015 constraints on $n_{\rm s}$ and $r$, is  $0.1 < \beta < 0.9$. 

As in the Starobinsky case, expressed by Eq.~(\ref{eq:approxStar}), we assume that $\sigma R \gg 1$. Therefore, Eq.(\ref{efoldseq}) can be approximated to:
\begin{eqnarray}
N &\approx & \frac{3}{4}\int_{\sigma R_f}^{\sigma R_i} d(\sigma R) \frac{\sigma R}{(1-\beta) (\sigma R)^{\beta + 1}} \; 
\sim \frac{3}{4}\frac{(\sigma R)^{1-\beta}}{(1 - \beta)^{2}}.\; \label{eq:N1} \quad
\end{eqnarray}
This latter allows us to express $\epsilon$, $\eta$ and $\zeta$ in terms of $N$ as 
\begin{eqnarray}
\epsilon &=& \frac{3}{4 (1 - \beta)^{2} N^{2}} \;, \\
\eta &=& -\frac{1}{N} \;,\\
\zeta &=& \frac{1}{(1-\beta)^{2}N^{2}}\;.
\end{eqnarray}
We choose $\beta = 1/2$ and $N=60$ as an explicit example in order to compare its predictions for the slow-roll parameters, and consequently the non-Gaussianity function, with those of the Starobinsky model. For this case the slow-roll parameters of the $\beta$-model are estimated as:
\begin{eqnarray}
\epsilon_{(\beta = 1/2)} &=& 8.3\e{-4} \;,\\
\eta_{(\beta = 1/2)} &=& -1.6\e{-2} \;,\\
\zeta_{(\beta = 1/2)} &=& 1.1\e{-3} \;,
\end{eqnarray}
from which we can compute the tensor-to-scalar ratio, scalar spectral index and running of the scalar index as:
\begin{equation}\label{rnsbeta}
r = 0.013\;, \quad n_{\rm s} = 0.966 \;, \quad \alpha_{\rm s} = -0.002\;.
\end{equation}

\begin{figure}[htbp]
	\centering
	\includegraphics[scale=0.5]{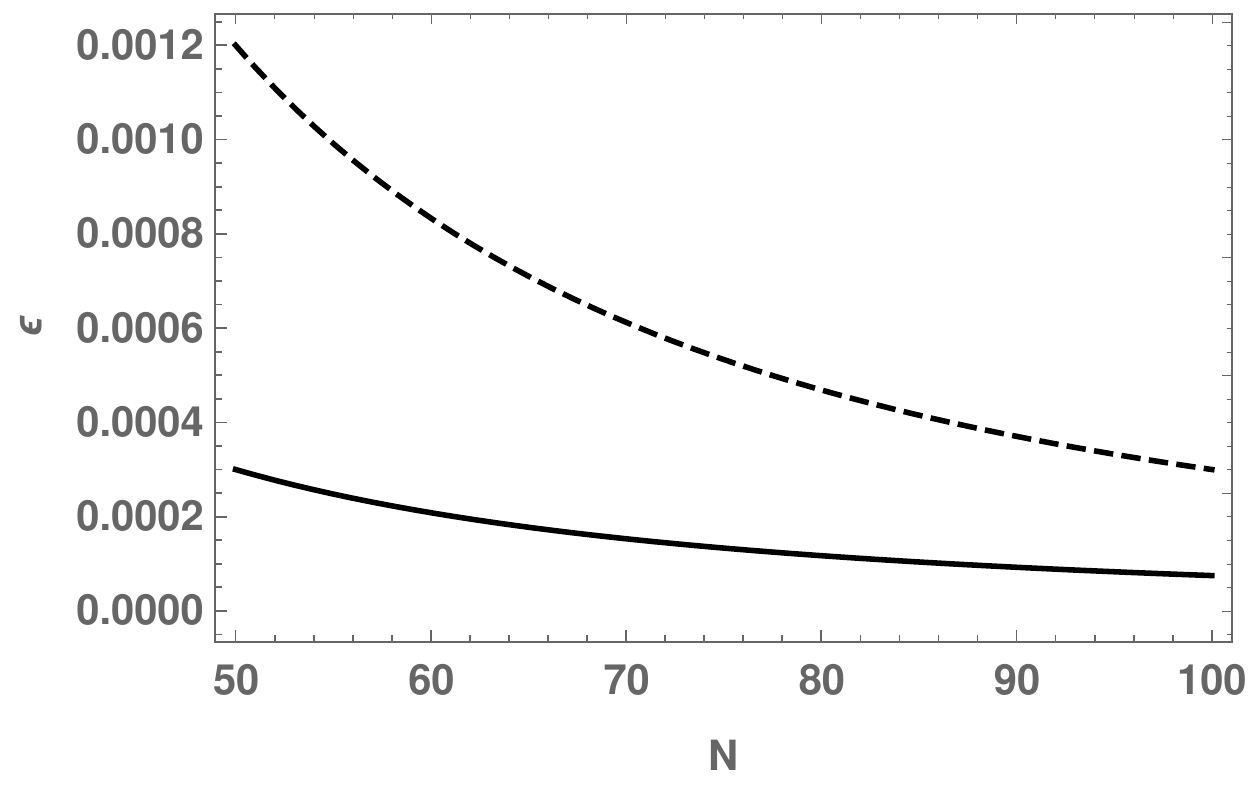}\includegraphics[scale=0.5]{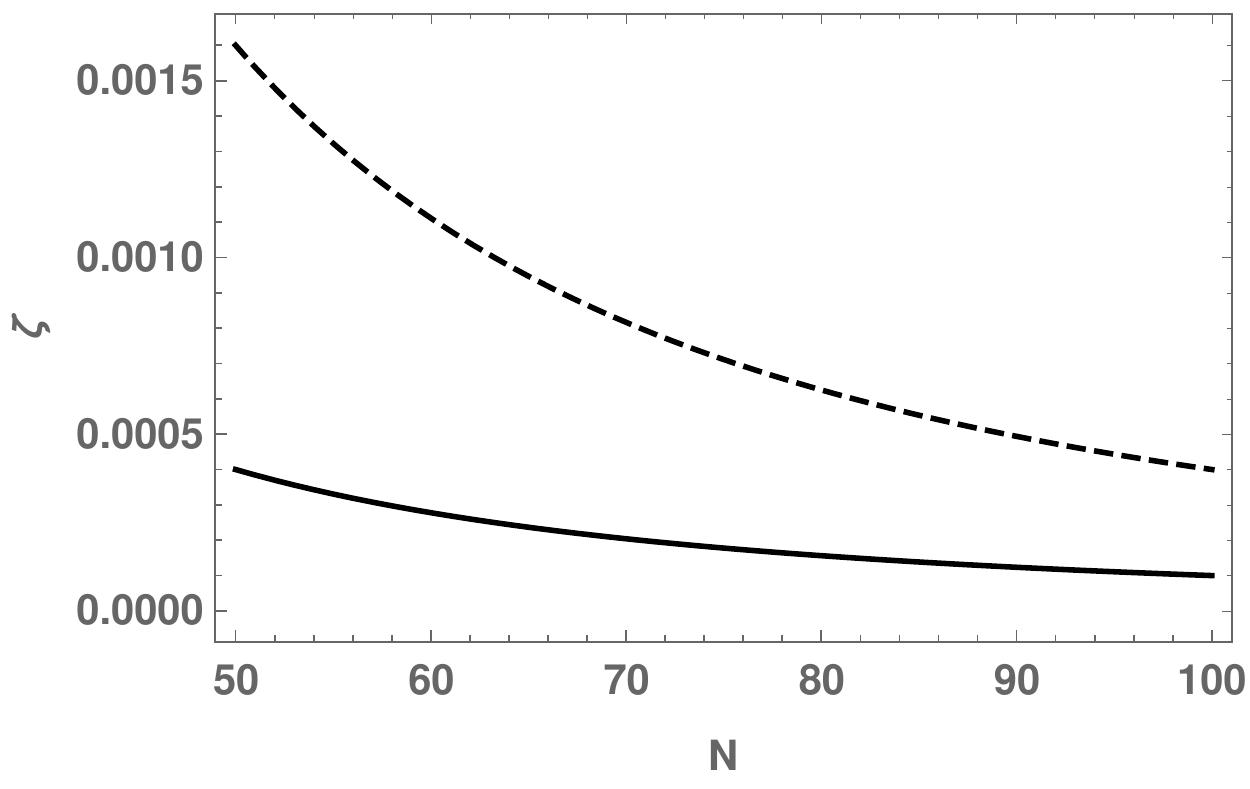}
	\caption{\textit{Left Panel.} Evolution of $\epsilon$ as function of $N$ for the Starobinsky model (solid line) and the $\beta$-model with $\beta = 1/2$ (dashed line). \textit{Right Panel.} Evolution of $\zeta$ as function of $N$ for the Starobinsky model (solid line) and the $\beta$-model with $\beta = 1/2$ (dashed line).}
	\label{fig:1}
    \end{figure}

\begin{figure}[htbp]
   	\centering
   	\includegraphics[scale=0.5]{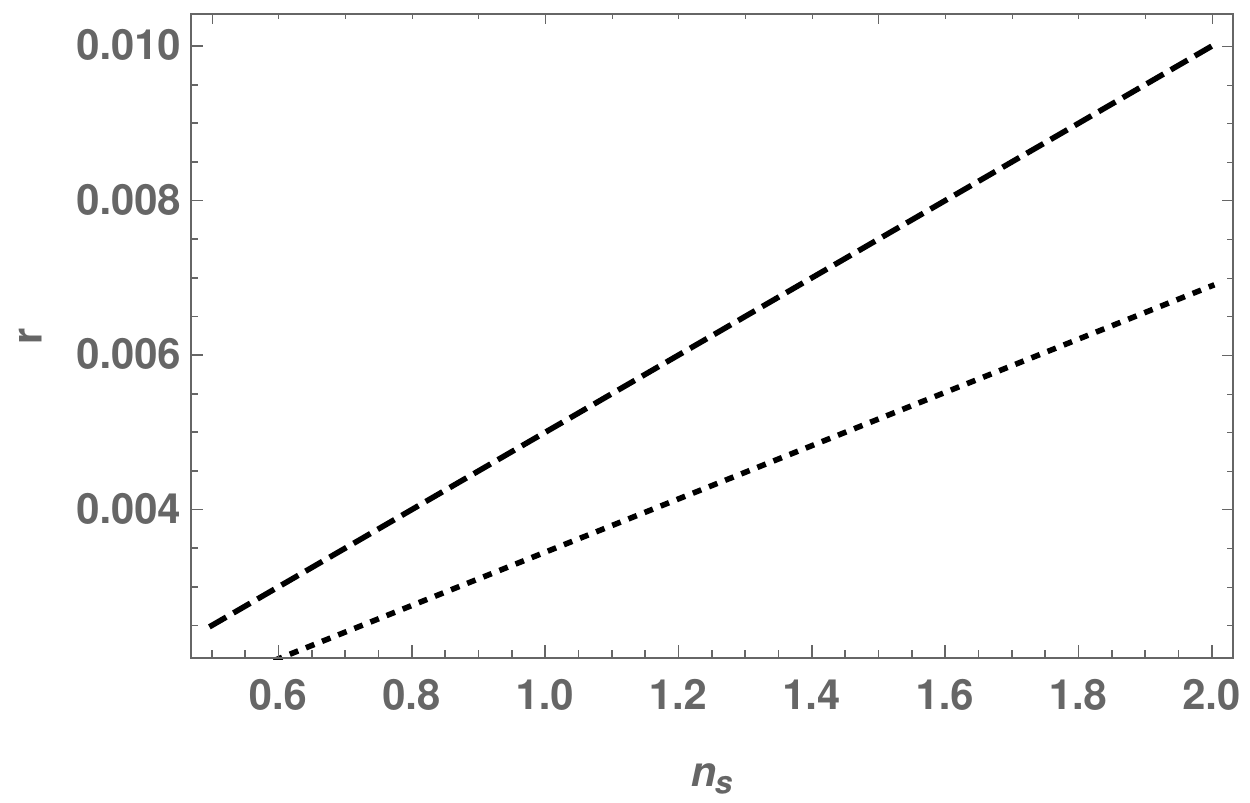}\includegraphics[scale=0.5]{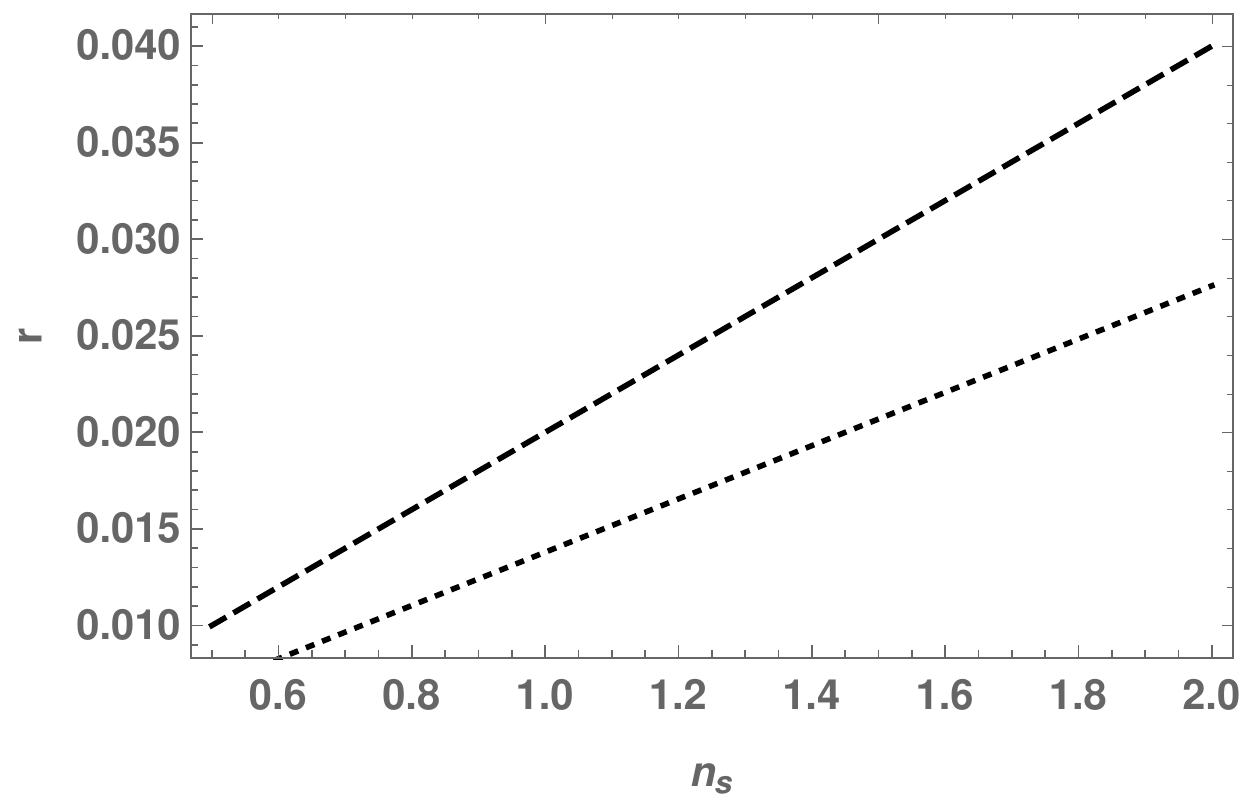}
   	\caption{\textit{Left Panel.} Prediction of the Starobinsky model for $N=50$ (dashed line) and $N=60$ (dotted line) on the space ($n_s,r$). \textit{Right Panel.} Prediction of the $\beta$-model for $N=50$ (dashed line) and $N=60$ (dotted line) on the space ($n_s,r$).}
   	\label{fig:2}
   \end{figure}

\begin{figure}[htbp]
      	\centering
      	\includegraphics[scale=0.5]{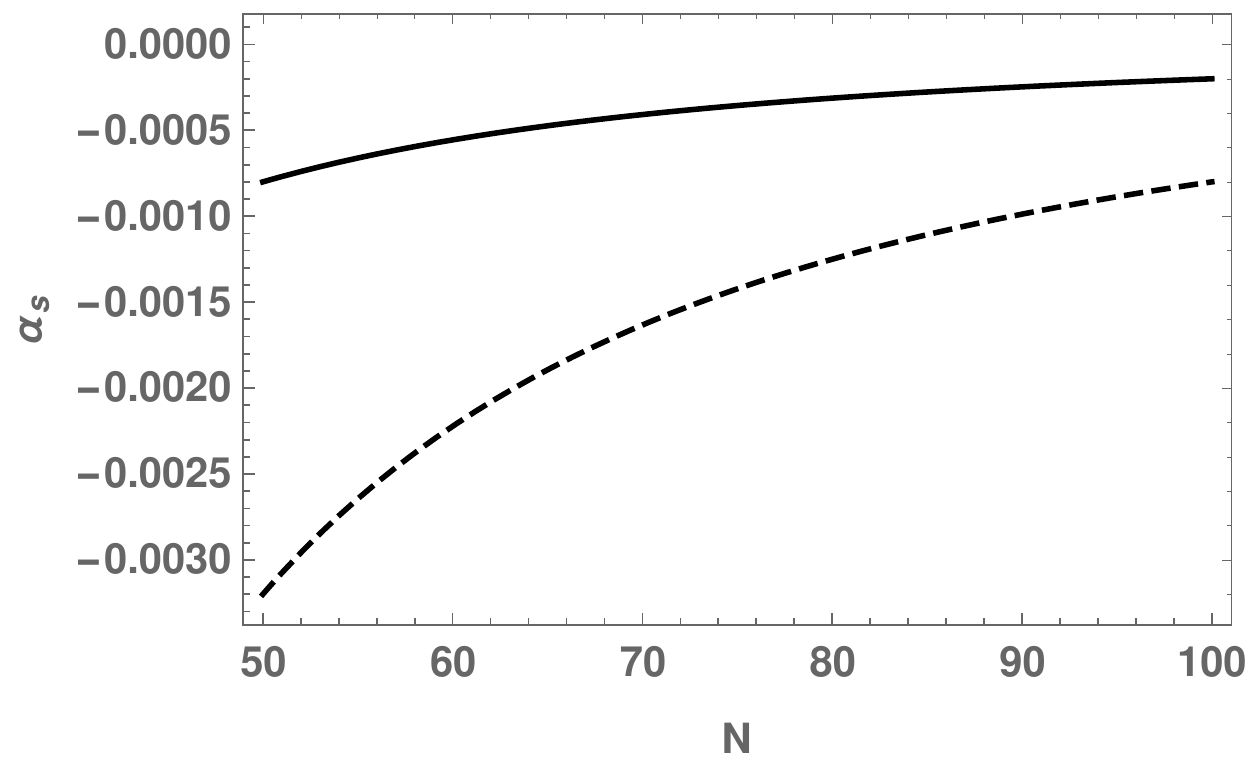}\includegraphics[scale=0.5]{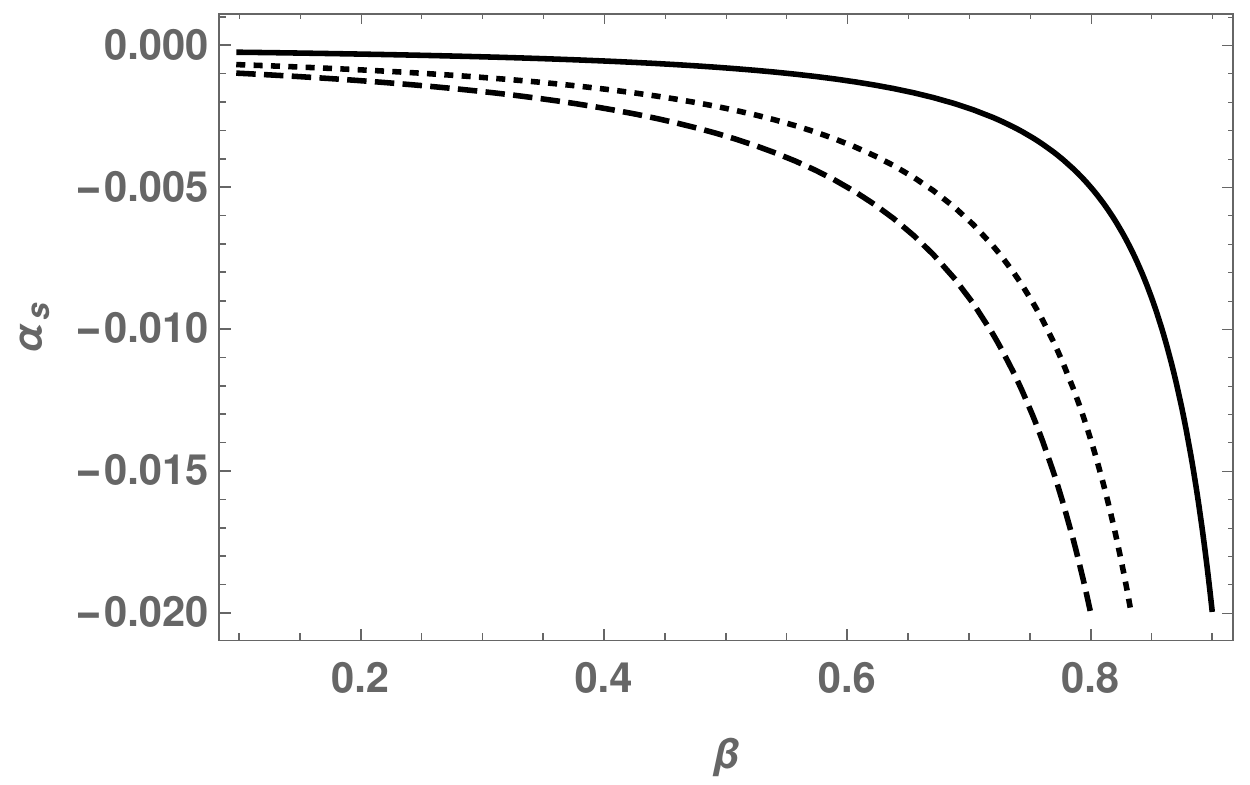}
      	\caption{\textit{Left Panel.} Evolution of $\alpha_{s}$ as function of $N$ for the Starobinsky model (solid line) and the $\beta$-model with $\beta = 1/2$ (dashed line). \textit{Right Panel.} Running of the spectral index for the $\beta$-model showing the cases : $N = 50$ (dashed line), $N = 60$ (dotted line) and $N = 100$ (solid line).}
      	\label{fig:3}
      \end{figure}

We show the evolution of the slow-roll parameters $\epsilon$ and $\zeta$ for the Starobinsky model and the $\beta$-model, with $\beta = 1/2$, as functions of $N$ in Figure~\ref{fig:1}. The evolution of the tensor-to-scalar ratio versus 
the scalar spectral index for the Starobinsky model and for the $\beta$-model, with $\beta = 1/2$, are shown in Figure~\ref{fig:2}. For completeness, we express $\alpha_{\rm s}$ for the Starobinsky model and the $\beta$-model, with $\beta = 1/2$ 
versus
$N$ and the evolution of $\alpha_{\rm s}$ \textit{vs} $\beta$ are presented in Figure~\ref{fig:3}.

	\section{Generic non-gaussianities for a $f(R)$ model}
	The slow-roll parameters defined in Eqs.~(\ref{eq:SR}) from the potential $U$ are related to the ones defined from the Hubble parameter as follows:
	\begin{eqnarray}\label{eqs:slowrolls}
	\epsilon_{U} = \epsilon &\approx& \epsilon_H \;, \\
	\eta_{U} = \eta &\approx& \epsilon_H +\eta_H ,\; \\
	\zeta_{U}= \zeta  &\approx& \zeta_H + 2\epsilon_H (3\eta_H - \epsilon_H). \;
	\end{eqnarray}
	Notice that for a successful inflation, 
	we need $\epsilon \ll 1$, $\eta \ll 1$ for a standard Starobinksy model and additionally $\zeta \ll 1$ for an $\alpha$-attractor model.
	Indirectly, large values of $\eta$ are likely to make $\epsilon$ and $\zeta$ grow as well.
	
	In standard inflation driven by a potential $U_{\chi}$ we have the equilateral non-linear parameter as \cite{DeFelice:2011zh}
	\begin{eqnarray}
	f^{\text{equil}}_{\text{NL}} &=&\frac{55}{36}\epsilon_H +\frac{5}{12}\eta_H \;,
	\end{eqnarray}
	therefore, using our definitions for the slow-roll parameters (\ref{eqs:slowrolls}), we can have a \textit{generic non-Gaussianity parameter for $f(R)$}:
	\begin{eqnarray}
	f^{\text{equil}}_{\text{NL}}= \frac{30}{27}\epsilon_U +\frac{5}{12}\eta_U \;.
	\end{eqnarray}
	From this \textit{generic} result, we can express the non-Gaussianity parameter in terms of $N$ for the Starobinsky model as:
	\begin{eqnarray}\label{eq:fnlstar}
	f^{\text{equil}}_{\text{NL}}= \frac{15}{18 N^{2}}-\frac{5}{12 N} \;.
	\end{eqnarray}
	and for the $\beta$-model:
	\begin{align}
	&f^{\text{equil}}_{\text{NL}} = \frac{15}{18(1-\beta)^{2}N^{2}}-\frac{5}{12N}\;.\label{eq:fnlb1}
	\end{align}
 
	When we choose $\beta = 1/2$, the non-Gaussianity parameter becomes:
	\begin{eqnarray}
	f^{\text{equil}}_{\text{NL}}= \frac{30}{9 N^{2}}-\frac{5}{12 N} \;.\label{eq:fnlbeta}
	\end{eqnarray}
	
\begin{figure}[htbp]
	  \centering
	  \includegraphics[scale=0.5]{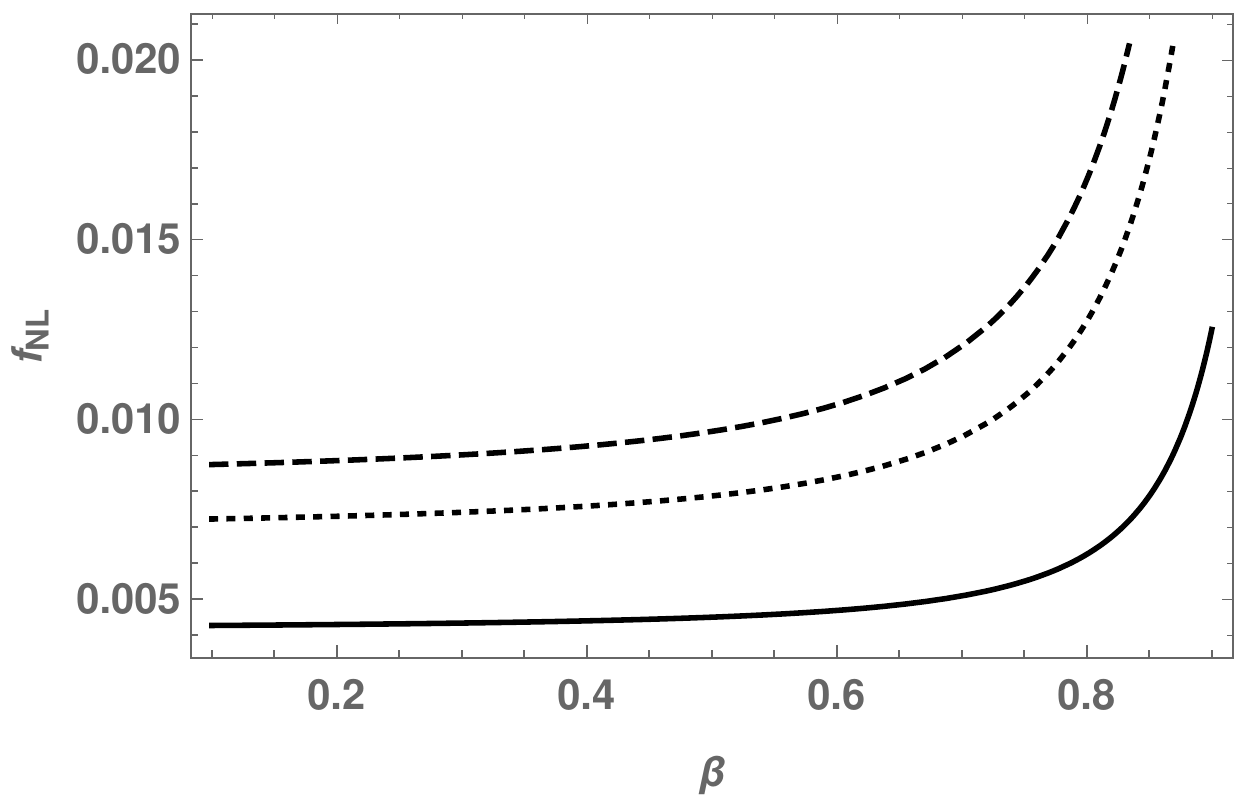}\includegraphics[scale=0.53]{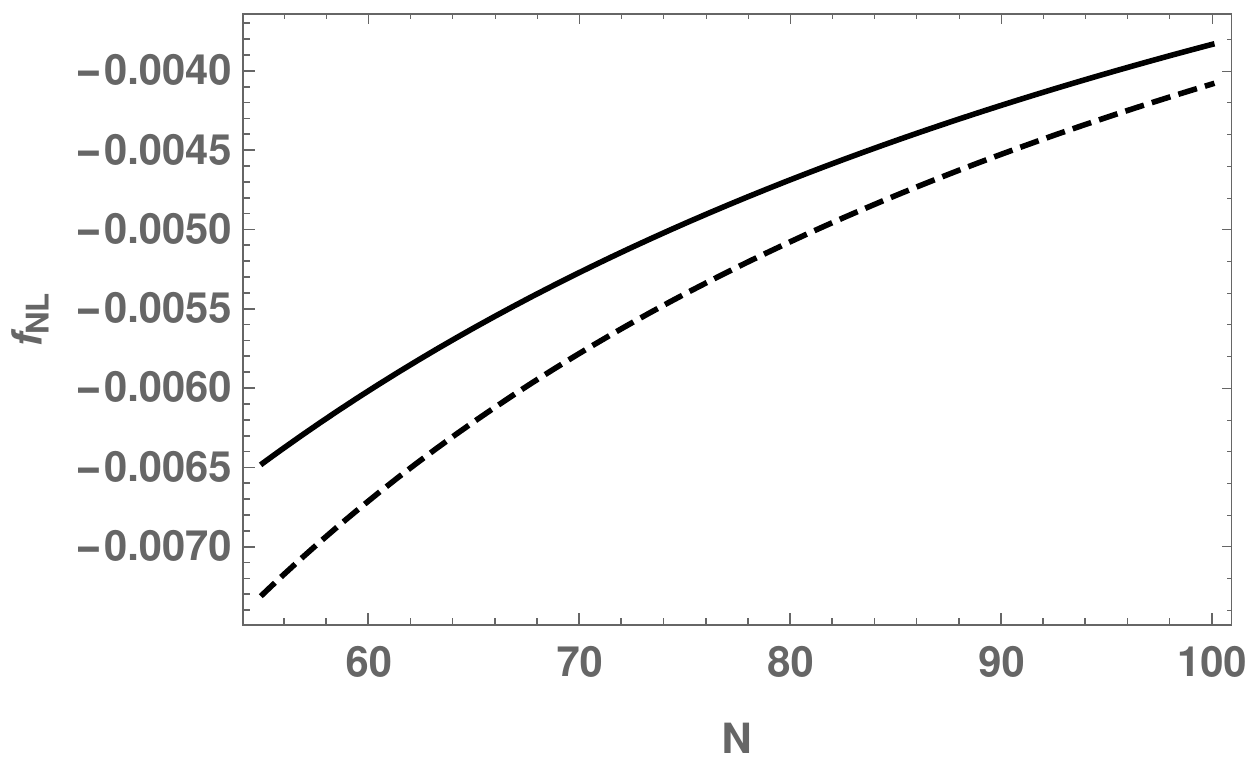}
	  \caption{\textit{Left Panel.} Non-gaussianity parameter for the $\beta$-model showing the cases : $N = 50$ (dashed line), $N = 60$ (dotted line) and $N = 100$ (solid line). \textit{Right Panel.} Non-gaussianity parameter evolution for the Starobinsky model (dashed line) and the $\beta$-model (solid line), with $\beta = 1/2$, as functions of $N$.}
	  \label{fig:4}
	  \end{figure}

     In Figure~\ref{fig:4} we present the behaviour of $f^{\text{equil}}_{\text{NL}}$, given by Eq.~(\ref{eq:fnlb1}). Also, we show the non-Gaussianity parameter for the Starobinsky model, given by Eq.~(\ref{eq:fnlstar}), and for the $\beta$-model with $\beta = 1/2$, given by Eq.~(\ref{eq:fnlbeta}), as functions of the number of e-folds.
	
	\section{Conclusions}
	In this paper we present \textit{formulae} for the inflationary slow-roll parameters ($\epsilon$, $\eta$ and $\zeta$) 
	and the non-Gaussianity function $f^{\text{equil}}_{\text{NL}}$ to be used in any kind of $f(R)$ proposals which satisfies the condition for slow-roll existence, \textit{i.e.}, any $f(R)$ that behaves like $f(R)/R^{2}\approx constant$.
	Also, as an example, we develop these quantities using 
	the Starobinsky model and the $\beta$-model, which is a $f(R)$ approximate reconstruction of the $\alpha$-Attractors class of inflationary models.
	As shown in Figure~\ref{fig:4}, these two models provide small non-Gaussianity parameters, which is in good agreement with recent constraints \cite{DeFelice:2011zh}, that despite the agreement of the Planck data sets with the Gaussian scenario, they do not rule out the presence of non-Gaussianities of small intensity.
	
	\acknowledgments
	C. E-R. is supported by ICTP Professor Visitor Program. This study was financed in part by the Coordena\c{c}\~{a}o de Aperfei\c{c}oamento de Pessoal de N\'{i}vel Superior - Brasil (CAPES) - Finance Code 001. Also it has been supported by CNPq (Brazil) and Fapes (Brazil). OFP thanks CAPES (Brazil) and the Humboldt foundation (Germany) for financial support and the Institut f\"ur Theoretische Physik of the Ruprecht-Karls-Universit\"at (Heidelberg) for kind hospitality. We thank professor David Wands for useful suggestions. Also we thank the anonymous Referee for their careful reading and their constructive comments.

\end{document}